\documentclass[%
reprint,
superscriptaddress,
 amsmath,amssymb,
 aps,
 prb,
]{revtex4-2}

\usepackage[colorlinks, linkcolor=blue, anchorcolor=blue, citecolor=blue, urlcolor=blue]{hyperref} 
\usepackage{graphicx}% Include figure files
\usepackage{dcolumn}% Align table columns on decimal point
\usepackage{bm}% bold math
\usepackage{float}
\usepackage{multirow}
\usepackage{setspace}
\begin{document}

\title{Unveiling resilient superconducting fluctuations in atomically thin NbSe$_2$ through Higgs mode spectroscopy}

\author{Yu~Du}
\thanks{These authors contributed equally to this work.}
\author{Gan~Liu}
\thanks{These authors contributed equally to this work.}
\author{Wei~Ruan}
\thanks{These authors contributed equally to this work.}
\author{Zhi~Fang}
\affiliation{National Laboratory of Solid State Microstructures and Department of Physics, Nanjing University, Nanjing 210093, China}

\author{Kenji Watanabe}
\affiliation{Research Center for Electronic and Optical Materials, National Institute for Materials Science, 1-1 Namiki, Tsukuba 305-0044, Japan}

\author{Takashi~Taniguchi}
\affiliation{Research Center for Materials Nanoarchitectonics, National Institute for Materials Science,  1-1 Namiki, Tsukuba 305-0044, Japan}

\author{Ronghua~Liu}
\affiliation{National Laboratory of Solid State Microstructures and Department of Physics, Nanjing University, Nanjing 210093, China}
\affiliation{Collaborative Innovation Center of Advanced Microstructures, Nanjing University, Nanjing 210093, China}

\author{Jian-Xin~Li}
\author{Xiaoxiang~Xi}
\email{xxi@nju.edu.cn}
\affiliation{National Laboratory of Solid State Microstructures and Department of Physics, Nanjing University, Nanjing 210093, China}
\affiliation{Collaborative Innovation Center of Advanced Microstructures, Nanjing University, Nanjing 210093, China}
\affiliation{Jiangsu Physical Science Research Center, Nanjing 210093, China}

\begin{abstract}
We report a combined electrical transport and optical study of the superconductivity in atomically thin NbSe$_2$. When subjected to an out-of-plane magnetic field, an anomalous metallic state emerges, characterized by a finite longitudinal resistance and a vanishing Hall resistance, suggesting the presence of particle-hole symmetry. We establish a superconducting Higgs mode in atomically thin samples, which reveals enduring superconducting fluctuations that withstand unexpectedly high reduced magnetic fields. These findings provide evidence of robust locally paired electrons in the anomalous metallic state, affirming its bosonic nature. 
\end{abstract}

%\date{\today}

\maketitle

The magnetic field ($H$)-temperature ($T$) phase diagram of type-II superconductors is dominated by a mixed state, which can retain zero dissipation as long as the vortices are pinned~\cite{Blatter1994}. The situation can be considerably different in the two-dimensional (2D) limit, as enhanced thermal and quantum fluctuations disrupt the superconducting order and induce dissipation. Notably, in many thin-film superconductors, a finite resistance much lower than the normal-state value has been observed under a perpendicular magnetic field, which persists down to the zero-temperature limit~\cite{Jaeger1989,Ephron1996,Wang2024}. The existence of this anomalous metallic state (AMS) contradicts the claim of the absence of 2D metallicity as proposed by the scaling theory of localization~\cite{Abrahams1979}. Research over the past decades has led to the notion that this state could be considered as a failed superconductor~\cite{Kapitulnik2019}, yet its origin remains unresolved~\cite{Feigelman1998,Spivak2001,Spivak2008,Shimshoni1998,Mason1999,Galitski2005,Das1999,Das2001,Dalidovich2002,Phillips2003,Mulligan2016}. 

Highly crystalline 2D superconductors are well suited for studying the AMS because of their exceptional cleanness~\cite{Saito2016}. In this low-disorder regime, a magnetic-field-induced superconductor-metal transition is typically seen~\cite{Yu2015,Tsen2016,Sajadi2018,Li2019,Liu2020,Xing2021}, and the low-field dissipative state is bound to be metallic. However, limited by the minute sample volume, experimental probes of the AMS in crystalline 2D superconductors have not advanced beyond dc transport, and new techniques are yet to be developed. Versatile probes are available for films with much larger size, revealing a particle-hole symmetry arising from uncondensed Cooper pairs based on vanishing Hall response~\cite{Breznay2017,Yang2019,Chen2021,Li2024b}, absence of cyclotron resonance measured by microwave spectroscopy~\cite{Wang2018}, and charge-2$e$ ($e$ is the elementary charge) quantum oscillation in nano-patterned films~\cite{Yang2019,Li2024b}. These results point to bosonic nature of the AMS, with local Cooper pairs dominating its electrodynamic response. 

Atomically thin NbSe$_2$ is a prototypical crystalline 2D superconductor in which an AMS has been debated. An initial study reported this phase occupying a large portion of the $H$-$T$ phase diagram, and the resistance obeying a power-law scaling was taken as the signature for a Bose metal~\cite{Tsen2016}, which arises due to the quantum phase fluctutations induced by the applied field~\cite{Das1999,Das2001,Dalidovich2002,Phillips2003}. However, further study showed that this AMS was eliminated by filtering the electromagnetic noise coupled to the sample in the transport measurement~\cite{Tamir2019}. Finite resistance resembling the AMS reemerged by increasing the driving current, which was attributed to vortex depinning~\cite{Tamir2019,Benyamini2019}. These findings reveal the fragility of the zero-resistance superconducting state in atomically thin NbSe$_2$ and raise doubts about whether the AMS encompasses any genuine physics.

In this letter, we conduct both electrical and optical measurements on NbSe$_2$ samples with varying thicknesses to examine the characteristics of the AMS resulting from fragile superconductivity. We found an extended magnetic field range in which the longitudinal resistance becomes finite whereas the Hall resistance vanishes, which occurs exclusively in atomically thin samples. We observed a thickness-dependent superconducting Higgs mode, a rarely seen collective mode that corresponds to the amplitude fluctuation of the superconducting order parameter~\cite{Sooryakumar1980,Measson2014,Grasset2018,Grasset2019}. Utilizing this sensitive probe, we reveal that superconducting fluctuations survive in the entire magnetic field range for the AMS. These results affirm the bosonic nature of the AMS and highlight its remarkable resilience to external magnetic fields.

\begin{figure}[t]
\centering
\includegraphics[width=\linewidth]{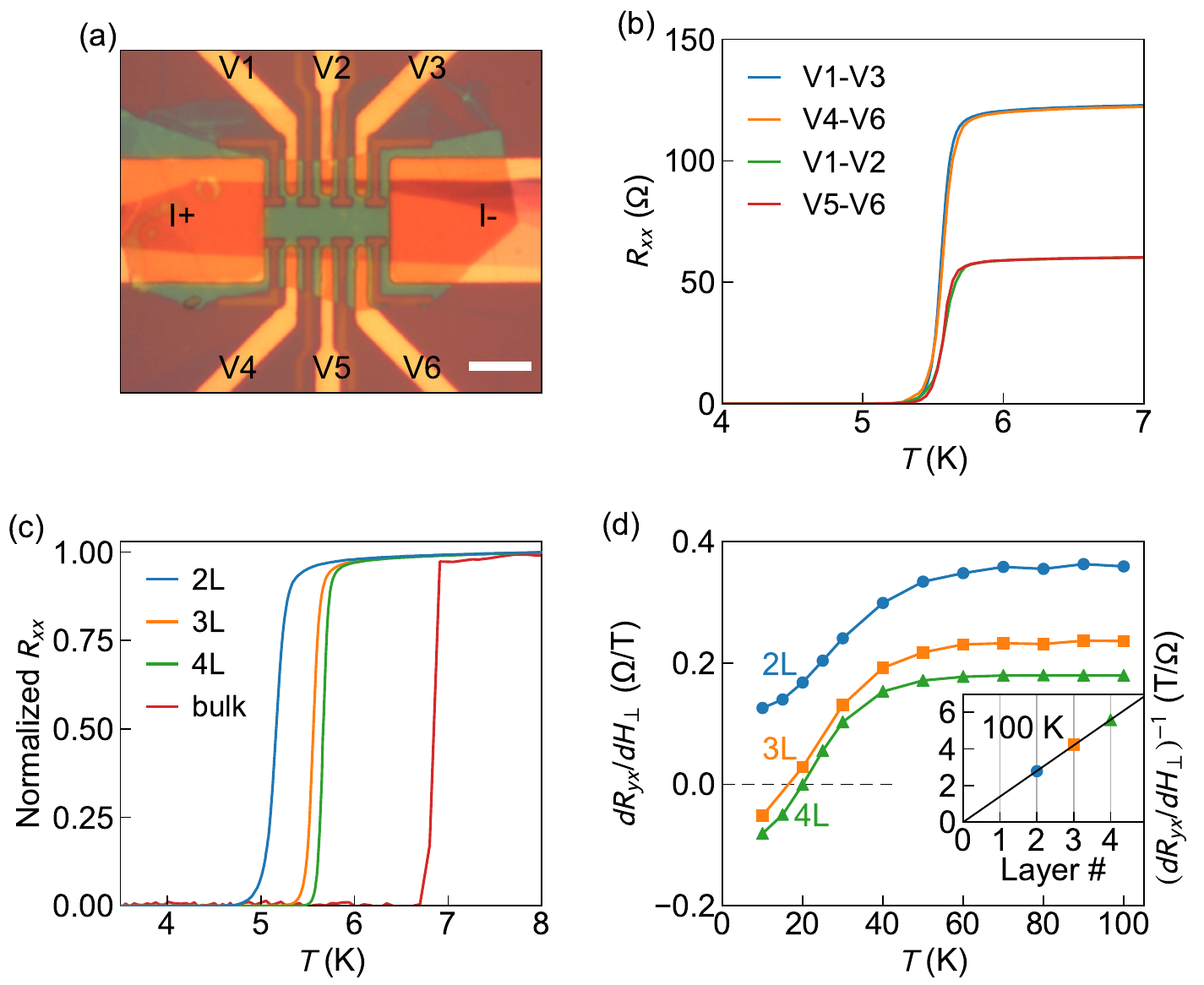}
\caption{(a) Optical image of a trilayer NbSe$_2$ device. The NbSe$_2$ transferred on Au electrodes and covered by h-BN is etched to form a regular Hall bar pattern. The I$+$ and I$-$ electrodes are used to source current and the rest are voltage probes. Scale bar: 10~$\mu$m. (b) Four-probe longitudinal resistance of the device shown in (a), measured using different combinations of voltage probes. (c) Temperature dependence of the normalized resistance for samples of differing thickness. (d) Temperature dependence of the slope in the magnetic field dependent Hall resistance for 2--4-layer samples. The inset shows the thickness dependence of the reciprocal of the data at 100~K fitted to a straight line with zero intercept.}
\label{Fig1}
\end{figure}

We developed a device fabrication method that both achieves a reliable Hall-bar geometry and preserves the intrinsic properties of atomically thin NbSe$_2$ (see details in Supplemental Note 1~\footnote{See Supplemental Material at [URL] for the methods for sample preparation, device fabrication, electrical transport and Raman measurements; $H$-$T$ phase diagram for other samples; analysis of the CDW amplitudon and Higgs mode; and discussion on the Berezinskii–Kosterlitz–Thouless transition.}). Figure~1(a) shows a typical device image. Longitudinal resistance $R_{xx}$ measured using different voltage probes yields consistent superconducting resistive transitions [Fig.~1(b)]. A monotonically reduced transition temperature ($T_c$, defined as the temperature at which the resistance drops to half of the normal-state value) with decreasing sample thickness is observed [Fig.~1(c)], as reported previously~\cite{Xi2016,Cao2015}. 

These devices facilitate precise characterization of the Hall resistance $R_{yx}$ in atomically thin NbSe$_2$. Figure~1(d) shows the slope in the magnetic-field dependence of $R_{yx}$ ($dR_{yx}/dH$) as a function of temperature. For samples of varying thickness, $dR_{yx}/dH$ remains nearly temperature independent above 60~K but decreases at low temperature due to the charge-density wave (CDW) transition~\cite{Huntley1974,Naito1982}. Since the Fermi surface of NbSe$_2$ comprises multiple hole pockets~\cite{Calandra2009}, the simple relation $(dR_{yx}/dH)^{-1}=ned$ ($n$ is the carrier density and $d$ is the film thickness) for a single-band metal is generally inapplicable. However, a linear thickness dependence was observed at 100~K [Fig.~1(d) inset], indicating that at this temperature, electronic scattering is isotropic~\cite{Ong1991}. The hole density, extracted to be $1.4\times 10^{22}$~cm$^{-3}$, is consistent with the bulk value~\cite{Naito1982} and remains constant regardless of sample thickness. This further suggests that the electronic band structure in NbSe$_2$ down to the bilayer thickness closely resembles that of the bulk material~\cite{Xu2018}.

\begin{figure}[t]
\centering
\includegraphics[width=\linewidth]{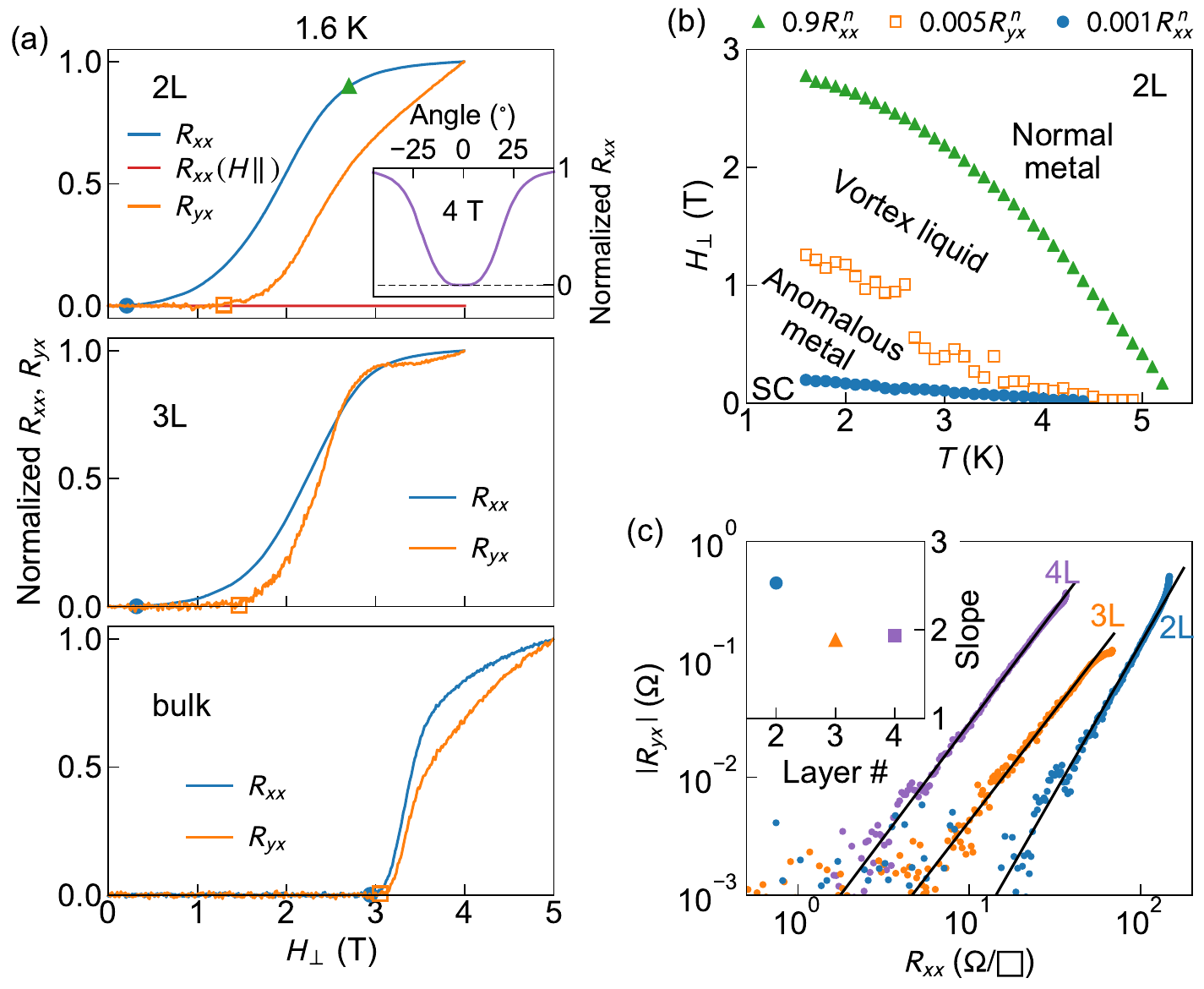}
\caption{(a) Magnetic-field dependence of $R_{xx}$ and $R_{yx}$ at 1.6~K for samples of varying thickness, normalized to the respective values at the maximum field shown. The circles, squares, and triangle mark $0.001R_{xx}^n$, $0.005R_{xy}^n$, and $0.9R_{xx}^n$, respectively. (b) $H$-$T$ phase diagram for bilayer NbSe$_2$. The phase boundaries are determined according to the legends. (c) Scaling showing the $|R_{yx}|\propto R_{xx}^{\alpha}$ relation, with the layer-number dependence of $\alpha$ (determined as the slope of fitted straight lines) plotted in the inset. All data are for perpendicular magnetic fields unless specified otherwise. The inset in the first panel in (a) shows the angle dependence of the normalized $R_{xx}$, with $0^{\circ}$ corresponding to the in-plane field. }
\label{Fig2}
\end{figure}

Focusing on the superconducting state, Fig.~2(a) compares the field dependence of $R_{xx}$ and $R_{yx}$ at 1.6~K for samples with different thickness. They turn on at different magnetic fields for bilayer and trilayer NbSe$_2$, whereas they become finite at nearly the same field in bulk NbSe$_2$. We define an upper bound for these critical fields according to $R_{xx}(H_{\mathrm{M1}})=0.001R_{xx}^n$ and $R_{yx}(H_{\mathrm{M2}})=0.005R_{yx}^n$, in which the superscript $n$ represents the normal state at 1.6~K and 4~T or 5~T. (These correspond to the noise floor of our measurements, see Supplemental Note 2.) They are marked in Fig.~2(a) as solid circles and open squares, respectively. Together with the perpendicular upper critical field $H_{c2}^{\perp}$, defined according to $R_{xx}(H_{c2}^{\perp})=0.9R_{xx}^n$, we establish the $H$-$T$ phase diagram for bilayer NbSe$_2$ [Fig.~2(b)]. The corresponding results for bulk, tetralayer, and trilayer samples are shown in Supplemental Fig.~4.

For $H_{\mathrm{M2}}<H<H_{c2}^{\perp}$, both $R_{xx}$ and $R_{yx}$ are finite, and they obey a scaling relation, $|R_{xy}|\propto R_{xx}^{\alpha}$, where $\alpha\sim 2$ regardless of the sample thickness [Fig.~2(c)]. Such a scaling was commonly observed in conventional and high-temperature superconductors for the vortex liquid regime~\cite{Okuma1997,Kang2002,Luo1992,Vinokur1993,Samoilov1993}. For $H_{\mathrm{M1}}<H<H_{\mathrm{M2}}$, the vanishing $R_{yx}$ in the presence of finite $R_{xx}$ has been proposed as indicative of the particle-hole symmetry within the AMS~\cite{Breznay2017}. Our data were collected without filters, and it has been demonstrated previously that the finite $R_{xx}$ disappears upon implementing filters~\cite{Tamir2019}. We stress that the sensitivity of the superconducting state reflects actual physics, based on the following observations. First, the fact that $R_{xx}$ and $R_{yx}$ exhibit different turn-on fields in bilayer and trilayer NbSe$_2$ but not in the bulk sample vindicates a fragile zero-$R_{xx}$ state only in atomically thin samples. Second, under an in-plane magnetic field, $R_{xx}$ remains zero up to 4~T for bilayer NbSe$_2$ [upper panel in Fig.~2(a)] because the orbital effect is essentially absent~\cite{Xi2016}. The field-orientation dependence of $R_{xx}$ [Fig.~2(a) inset] proves that the AMS is associated with the orbital effect. Third, if electromagnetic noise were responsible for the finite $R_{xx}$ by driving vortex motion, $R_{xx}$ and $R_{yx}$ should have turned on simultaneously according to the models of Bardeen and Stephen~\cite{Bardeen1965} or Nozi\`{e}res and Vinen~\cite{Nozieres1966}. Taken together, our results suggest that the AMS is a manifestation of the fragile superconductivity in atomically thin NbSe$_2$.

\begin{figure}[t]
\centering
\includegraphics[width=\linewidth]{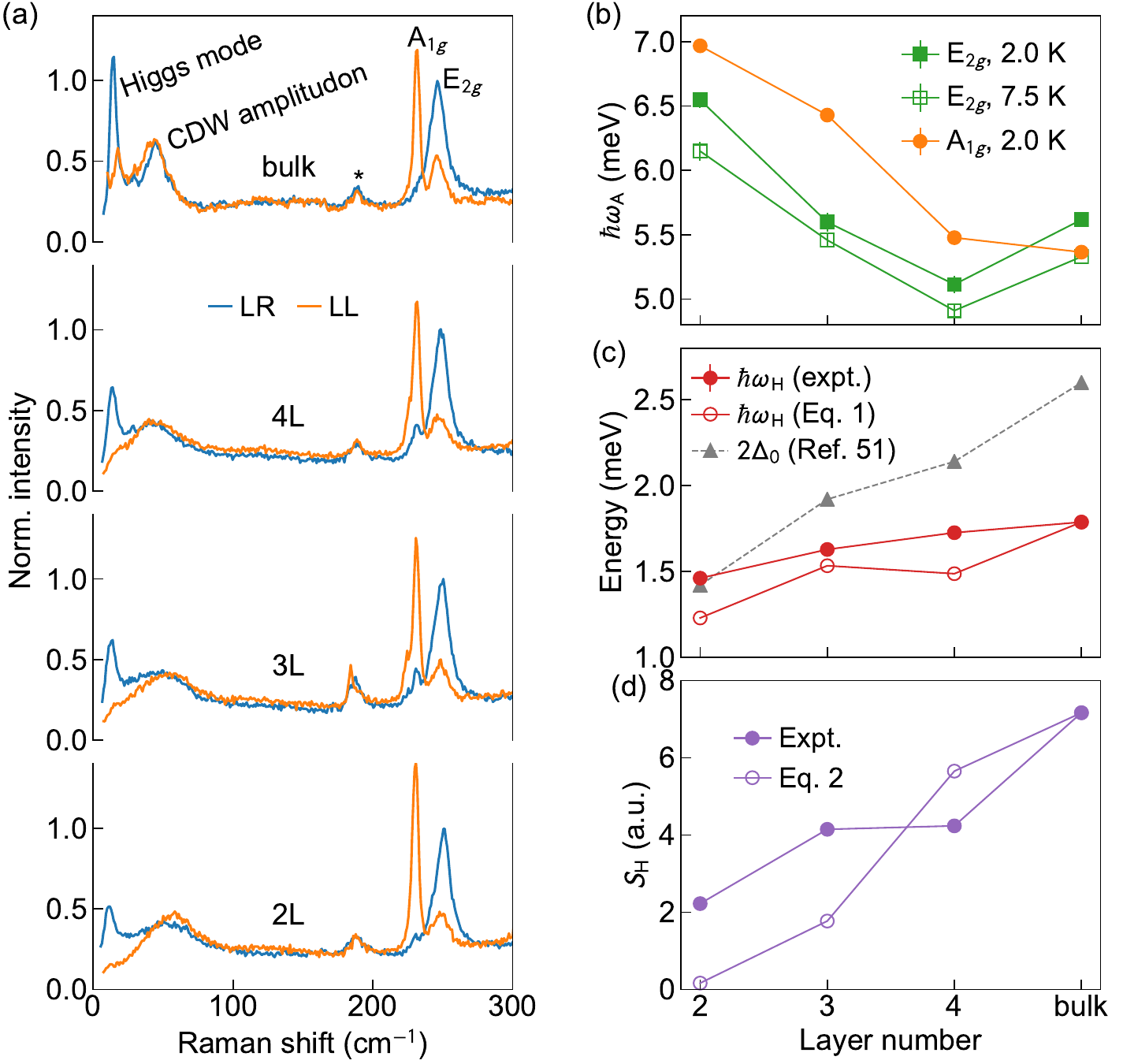}
\caption{(a) Raman spectra for NbSe$_2$ samples with varying thickness measured at 2.0~K in the LR and LL polarization configurations. The spectra have been normalized to the corresponding $E_{2g}$ phonon intensity because it is almost thickness independent~\cite{Lin2022}. (b)--(d) Thickness dependence of the CDW amplitudon energy (b), the Higgs mode energy and the $2\Delta_0$ values from Ref.~\cite{Khestanova2018} (c), and the spectral weight for the Higgs mode (d). Error bars are derived from the analysis.}
\label{Fig3}
\end{figure}

\begin{figure*}[t]
\centering
\includegraphics[width=0.8\linewidth]{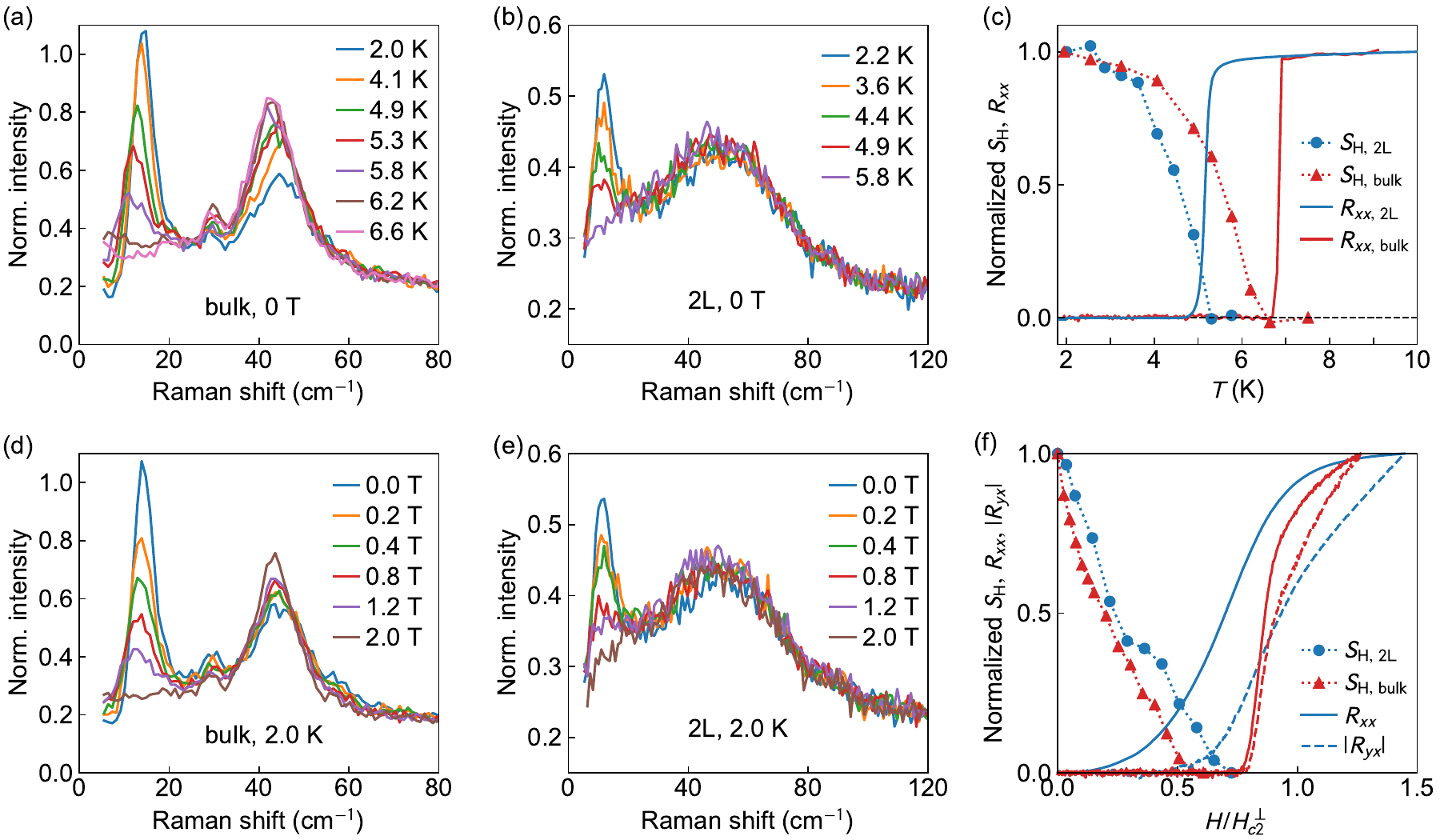}
\caption{(a) and (b) Temperature dependence of the Raman spectra for bulk (a) and bilayer (b) NbSe$_2$ measured at 0 T. (c) Temperature dependence of the normalized Higgs mode spectral weight (symbols) and $R_{xx}$ (solid lines) for bilayer (blue) and bulk (red) NbSe$_2$. (d) and (e) Perpendicular magnetic field dependence of the Raman spectra for bulk (d) and bilayer (e) NbSe$_2$ measured at 2.0~K. (f) Reduced field dependence of the normalized Higgs mode spectral weight (symbols), $R_{xx}$ (solid lines), and $|R_{yx}|$ (dashed lines) for bilayer (blue) and bulk (red) NbSe$_2$. Error bars in (c) and (f) are derived from the analysis.}
\label{Fig4}
\end{figure*}

To further probe this unusual state, we performed Higgs mode spectroscopy on these samples. We start from a bulk-like flake with a $T_c$ of 6.8~K. Figure~3(a) shows its Raman spectra at 2.0~K measured in the LL and LR polarization configurations, which detect the $A_{1g}$ and $E_{2g}$ channels, respectively. Two main lattice modes appear above 200~cm$^{-1}$ and a weak CDW-induced zone-folded mode~\cite{Lin2020} is marked by the asterisk. Focusing on the low-wavenumber region, two nearly-degenerate CDW amplitudons~\cite{Lin2020} are accompanied by Higgs modes near 20 cm$^{-1}$. The Higgs modes in the Raman response of bulk NbSe$_2$ are established signatures of superconductivity, which are activated through their coupling with the CDW amplitudon~\cite{Sooryakumar1980,Measson2014}. Intuitively, the Raman-active CDW amplitudon modulates the electronic states at the Fermi level and shakes the superconducting condensate, and this electron-phonon coupling mechanism makes the Higgs mode also visible in the Raman response~\cite{Littlewood1981,Cea2014}. 

As the sample thickness decreases, the Higgs mode persists down to bilayer NbSe$_2$ [Fig.~3(a)]. While the CDW amplitudons in the $E_{2g}$ and $A_{1g}$ channels are almost degenerate and comparable in intensity, the Higgs mode shows up predominantly in the $E_{2g}$ channel. Fitting analysis (see Supplemental Note 3~\cite{Note1}) yields the mode parameters shown in Fig.~3(b)--(d). The increased amplitudon energy $\hbar\omega_{\mathrm{A}}$ ($\hbar$ is the Planck constant) in atomically thin samples corresponds to an enhanced CDW order~\cite{Xi2015}. The non-monotonic thickness dependence of the $E_{2g}$ CDW amplitudon energy requires further work to clarify. The Higgs mode energy $\hbar\omega_{\mathrm{H}}$ decreases as the thickness is reduced, consistent with the suppression of the superconducting gap $2\Delta_0$ obtained from tunneling measurements~\cite{Khestanova2018} [Fig.~3(c)], and its spectral weight decreases accordingly [Fig.~3(d)]. For all samples, the coupling between the CDW amplitudon and the superconducting Higgs mode pushes the former to higher energy and also broadens it [see Fig.~4(a)--(b) and the open and filled squares in Fig.~3(b)]. Both effects are due to the resonance between these modes~\cite{Littlewood1981,Cea2014}.

We compare the thickness dependence of the Higgs mode energy $\hbar\omega_{\mathrm{H}}$ and spectral weight $S_{\mathrm{H}}$ with the theory put forward by Littlewood and Varma~\cite{Littlewood1981}, which predicts that 
\begin{align}
    \hbar\omega_{\mathrm{H}} &= 2\Delta_0\left[1-\frac{2a^2\delta^4}{\pi(1-\delta^2)\lambda^8}\right],\\
 S_{\mathrm{H}}&=\frac{8a^2\delta^5}{\pi^2(1-\delta^2)^3\lambda^8}.
\end{align}
Here $\delta=2\Delta_0/(\hbar\omega_{\mathrm{A}})$ and $\hbar\omega_{\mathrm{A}}$ is the normal-state value at 7.5~K. $\lambda$ is the BCS coupling constant, which is estimated according to $k_{\mathrm{B}}T_c=1.13E_{\mathrm{D}}e^{-1/\lambda}$, where $k_{\mathrm{B}}$ is the Boltzmann constant and $E_{\mathrm{D}}$ is the Debye cutoff energy~\cite{Harper1977}. The coefficient $a$ is mainly determined by the density of states at the Fermi level and its derivative with respect to the CDW lattice distortion, which we assume to be independent of thickness. It is determined from Eq.~(1) using the bulk parameters, which in turn yields the thickness dependence of $\hbar\omega_{\mathrm{H}}$ and $S_{\mathrm{H}}$ according to Eqs. (1) and (2) based on the thickness dependent $\Delta_0$ and $T_c$~\cite{Khestanova2018}. The calculated $S_{\mathrm{H}}$ has been normalized to the experimental value for the bulk sample. The theory captures the observed thickness dependence of $\omega_{\mathrm{H}}$ [Fig.~3(c)], but large discrepancy is seen for $S_{\mathrm{H}}$ [Fig.~3(d)], as $\delta$ is substantially suppressed in atomically thin NbSe$_2$ and the strong $\delta$ dependence in Eq.~(2) amplifies this effect. A smaller $\delta$ also implies that the resonance between the Higgs mode and the CDW amplitudon is suppressed, which leads to a smaller renormalization of $\hbar\omega_{\mathrm{H}}$ with respect to $2\Delta_0$ [Fig.~3(c)].

Having established the Higgs modes in atomically thin NbSe$_2$, we now use it to probe the superconducting properties. Figure~4(a) and 4(b) show the temperature dependent Raman spectra for bulk and bilayer NbSe$_2$, respectively, measured at zero field. The $S_{\mathrm{H}}$ in both samples decreases systematically upon warming towards $T_c$ and disappears when $R_{xx}$ turns finite [Fig.~4(c), see analysis method in Supplemental Note 3~\cite{Note1}], consistent with the Higgs mode being a characterization of the superconducting pairing amplitude. The phase fluctuation regime above the Berezinskii–Kosterlitz–Thouless transition temperature in bilayer NbSe$_2$ is difficult to access using the Higgs mode mainly because at this temperature the mode has become too weak to detect~\cite{Note1,Benfatto2009}.

We next examine how the Higgs mode evolves under an out-of-plane magnetic field. Figure~4(d) shows the Raman spectra of bulk NbSe$_2$ at selected magnetic fields and 2.0~K. The magnetic field quenches the Higgs mode, and the spectral change is similar to the effect of raising temperature. However, unlike the persistence of the Higgs mode up to $T_c$ under a zero field, it is no longer observable above 2~T, which is about $0.5H_{c2}^{\perp}$. Similar results were reported previously~\cite{Sooryakumar1981}. Strikingly, although bilayer NbSe$_2$ has a smaller $H_{c2}^{\perp}$, its Higgs mode experiences a field-induced quenching at an even slower rate especially below 0.5~T [Fig.~4(e)]. On the scale of the reduced field ($H/H_{c2}^{\perp}$), this effect becomes more apparent [Fig.~4(f)]. In fact, in bilayer NbSe$_2$, $S_{\mathrm{H}}$ persists until $R_{yx}$ turns finite, meaning that the Higgs mode is visible in the entire AMS. Trilayer and tetralayer samples show a similar behavior (see Supplemental Fig.~9).

These results suggest different nature of the vortex state in bulk and atomically thin NbSe$_2$. For BCS superconductors, pair breaking induced by magnetic impurities has been theoretically shown to strongly suppress the Higgs mode amplitude before the gapless state is reached~\cite{Li2024}. The effect of magnetic-field-induced pair breaking is expected to be similar~\cite{Maki1969}, which, combined with the spatial inhomogeneity of the superconducting order in the mixed state, may account for the suppression of the $S_{\mathrm{H}}$ in bulk NbSe$_2$. A previous study also interpreted the field dependence along these lines, though the critical field for the complete suppression of the Higgs mode remains unresolved~\cite{Sooryakumar1981}. For the AMS in bilayer NbSe$_2$, a microscopic understanding of how a magnetic field affects the superconductivity is lacking. The classical pair-breaking theory~\cite{Maki1969} does not account for fluctuation effects, making it insufficient to fully explain our results in 2D NbSe$_2$. Based on our results, the finite $R_{xx}$ shows the loss of global superconductivity, but the vanishing $R_{yx}$ and finite $S_{\mathrm{H}}$ indicate superconducting correlations. This result is similar to the AMS in InO$_x$ films, which, according to microwave spectroscopy, exhibits superconducting response only on short length and time scales~\cite{Liu2013}. One may argue that the resilience of the Higgs mode in bilayer NbSe$_2$ may be associated with the enhanced CDW order. However, this scenario is unlikely given the strong suppression of the superconducting gap and the decreased coupling of the Higgs mode and the CDW amplitudon at this thickness.

Our results suggest that the AMS is characterized by fluctuating local pairing, which fails to condense. Theories focusing on the role of phase fluctuation in disrupting global superconductivity could provide valuable understanding of the phenomena observed~\cite{Das1999,Das2001,Dalidovich2002,Phillips2003}. The vanishing Hall resistance with particle-hole symmetry cannot be reconciled with electrical transport involving a percolating normal-metal component, because it will dominate the Hall response and render $R_{yx}$ finite~\cite{Breznay2017}. This imposes limitations on theories centered around superconducting puddles within a metal matrix~\cite{Spivak2001,Spivak2008} or dissipation involving fermionic degrees of freedom~\cite{Shimshoni1998,Mason1999,Galitski2005}. 

In conclusion, we have observed vanishing Hall response in the presence of dissipation in atomically thin NbSe$_2$, which evidences an AMS with particle-hole symmetry. The superconducting Higgs mode persists in atomically thin samples and serves as a sensitive probe of superconducting fluctuation. These measurements uncover that the AMS originates from a fragile superconducting state, yet demonstrates resilience by maintaining fluctuating Cooper pairs up to a high reduced magnetic field. Our results offer new insight into the anomalous vortex matter in 2D crystalline superconductors and call for future studies on its microscopic structure.

We thank Haiwen Liu for helpful discussions and Dongjing Lin and Tianyu Qiu for assistance with device fabrication. This work was supported by the National Key Research and Development Program of China (Grant No.~2024YFA1409100), the Natural Science Foundation of Jiangsu Province (Grant Nos. BK20231529 and BK20233001), the Fundamental Research Funds for the Central Universities (Grant No. 0204-14380233), the National Natural Science Foundation of China (Grant Nos. 12474170, 11774151, and 123B2059), and the National Postdoctoral Program for Innovative Talents (Grant No. BX20240160). K.W. and T.T. acknowledge support from the JSPS KAKENHI (Grant Nos. 20H00354 and 23H02052) and World Premier International Research Center Initiative (WPI), MEXT, Japan.

%\bibliography{reference}% Produces the bibliography via BibTeX.


\end{document}